\documentclass[preprint,showpacs]{revtex4}  
\usepackage{epsfig}  
 \def\r#1{(\ref{#1})}
 
\begin{document}  
\title{Position and Momentum Entanglement  
 of Dipole-Dipole Interacting Atoms in Optical  
Lattices: The Einstein-Podolsky-Rosen Paradox on a Lattice}  
\author{Tom\'{a}\v{s} Opatrn\'{y}$^{1}$, 
Michal Kol\'{a}\v{r}$^{1}$ 
Gershon Kurizki$^{2}$,
and
Bimalendu Deb$^{3}$}  
\address{  
$^{1}$ Department of Theoretical Physics, Palack\'{y} University,  
77146 Olomouc, Czech Republic \\  
$^{2}$Department of Chemical Physics, Weizmann Institute of Science,  
76100 Rehovot, Israel \\   
$^{3}$  Physical Research Laboratory,  
Ahmedabad-380009, India \\  
}  
\date{\today}  
  
\begin{abstract}  
We study a possible realization of the position- and    
momentum-correlated
atomic   pairs that are confined to adjacent sites of two mutually shifted
optical lattices   and are entangled via laser-induced dipole-dipole
interactions.     The Einstein-Podolsky-Rosen (EPR) ``paradox''  
[Phys. Rev. {\bf 47,} 777   
(1935)] with translational variables is then
modified by   lattice-diffraction effects.
This ``paradox'' can be verified to a high degree
of   accuracy in this scheme.  
\end{abstract}  
  
\maketitle  
   
\section{INTRODUCTION}\label{sec:1}

Einstein, Podolsky and Rosen (EPR) \cite{EPR} put forth the question of
whether  the quantum mechanical description of physical reality is
complete, giving the example of a two-particle quantum state showing peculiar
correlations (dubbed ``entanglement'' or
``Verschr\"{a}nkung'' by Schr\"{o}dinger \cite{sch}): if 
one measures the position or momentum of one
particle, one can predict with certainty the outcome of 
measuring their counterpart for the other particle.  Thus,
depending on which measurement is chosen for the first particle, the value of 
either the position or momentum can be predicted with arbitrary
precision for the second particle.
The ensuing controversy has revolved around the interpretation of the
EPR problem and its implications on quantum theory \cite{Bohr}. Later, Bohm
considered \cite{Boh} two entangled spin-1/2 particles, which have
become the focus of attention on this EPR issue: their {\em
discrete-variable} entangled states have served to demonstrate the
incompatibility of quantum mechanics with local realism, by the
violation of Bell's inequality 
\cite{Bell64,CHSH69,Aspect82a,Aspect82,Perrie85,OuMandel,Tittel,Weihs,Rowe,Rauch}. 
In recent years, there has been revival of interest in {\em 
continuous-variable} entanglement, in the spirit of the original EPR
problem
\cite{Gisin91,GisinPeres92,Reid,Ou,Silberhorn01,BraunsteinKimble,Furusawa98,Silberhorn02,Polzik,PolzikNature,LloydSlotine98,Braunstein98,Braunstein98N,LloydBraunstein99,Parkins,ZhangBraunstein,opa01,PRL03}.

\begin{figure}[t!]  
\centerline{\epsfig{figure=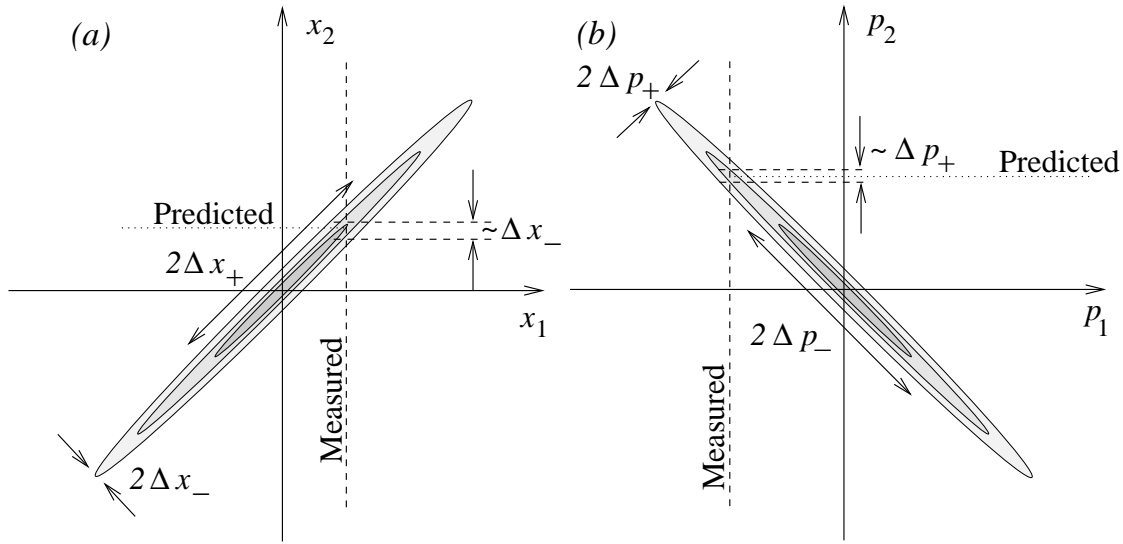,width=0.9\linewidth}}  
\caption{  
Joint probability distribution of positions $(a)$ and of momenta $(b)$
of EPR-pair ensembles. If one
measures the position of particle 1, one can predict 
the position of particle 2 with
uncertainty $\approx \Delta x_-$, whereas if one
measures momentum of particle 1, one can predict 
the momentum of particle 2 with
uncertainty $\approx \Delta p_+$.  
\label{figEPR}  
}   
\end{figure}

The original {\em ideal} EPR  \cite{EPR} state of two particles---1  
and 2, is, respectively, represented as follows in their coordinates 
or momenta (in one  
dimension),  
\begin{eqnarray}  
 \langle x_1, x_2| \psi_{\rm EPR} \rangle   
 &=& \delta (x_1 - x_2), \nonumber \\  
 \langle p_1, p_2| \psi_{\rm EPR} \rangle   
 &=& \delta (p_1 + p_2) . 
\label{EPRstateDelta} 
\end{eqnarray}  
If two particles are prepared in such a state, and one measures the value 
of $x_1$ (or  
$p_1$) of particle 1, one can predict the result of measuring  $x_2$   
 (or $p_2$,  
respectively) with perfect precision. 
The state of Eq. (\ref{EPRstateDelta}) would, 
however, occupy infinite space and have 
infinite kinetic energy.
One can consider more realistic variants of this state, e.g., a
Gaussian state given by
\begin{eqnarray}  
 \langle x_1, x_2| \psi_{\rm EPR} \rangle   
 &=& \frac{1}{\sqrt{\pi \Delta x_- \Delta x_+}}
 \exp \left( - \frac{(x_1-x_2)^2}{4 \Delta x_-^2}  \right)
 \exp \left( - \frac{(x_1+x_2)^2}{4 \Delta x_+^2}  \right)
 , \nonumber \\  
 \langle p_1, p_2| \psi_{\rm EPR} \rangle   
 &=& \frac{1}{\sqrt{\pi \Delta p_- \Delta p_+}}
 \exp \left( - \frac{(p_1-p_2)^2}{4 \Delta p_-^2}  \right)
 \exp \left( - \frac{(p_1+p_2)^2}{4 \Delta p_+^2}  \right) ,
\label{EPRstateGauss}
\end{eqnarray}  
where $\Delta p_{\pm} \equiv \hbar/\Delta x_{\pm}$ (see Fig. \ref{figEPR}). 
The original EPR state corresponds to
the limit $\Delta x_-/\Delta x_+ \to 0$. After
measuring the position  of particle 1, the position of 
particle 2 is centered at
\begin{eqnarray}
\bar x_2 = x_1 \frac{1-\left(\frac{\Delta x_-}{\Delta x_+}\right)^2}
{1+\left(\frac{\Delta x_-}{\Delta x_+}\right)^2} ,
\end{eqnarray}
with the uncertainty $\Delta x_-\left[ 1 + \left(\frac{\Delta x_-} {\Delta x_+}
\right) ^2 \right]^{-1/2}$. In the limit of $\Delta x_-/\Delta x_+  \ll 1$, the
position of particle 2 is centered at $x_1$, and its uncertainty
is $\approx \Delta x_-$. Similar relations hold also for the
momentum of particle 2
after the momentum of particle 1 is measured. Thus, either of the two
conjugate quantities of particle 2 can be predicted with arbitrarily 
high precision. Of course, the Heisenberg uncertainty relation is 
not violated,
since for a single system one can measure only one of the two
conjugate quantities.

Approximate versions of the translational EPR state, wherein the  
$\delta$-function correlations are replaced by finite-width distributions,
have   been shown to characterize the quadratures of the two optical-field
outputs of   parametric downconversion \cite{Reid,Ou}, or of a fiber
interferometer with Kerr nonlinearity \cite{Silberhorn01}. Such states allow
for various schemes of continuous-variable quantum information processing such
as quantum  teleportation \cite{BraunsteinKimble,Furusawa98} or quantum
cryptography \cite{Silberhorn02}.  A similar state has also been predicted and
realized using collective spins of large atomic samples
\cite{Polzik,PolzikNature}. It has been shown that if suitable interaction
schemes can be realized,  continuous-variable quantum states of the original
EPR type could even serve for quantum computation
\cite{Braunstein98,Braunstein98N,LloydSlotine98,LloydBraunstein99}.

Notwithstanding its applications to quantum information processing,
 the translational EPR state of
Eq. (\ref{EPRstateGauss}) does not entail
a violation of local realism: such a state has a {\em non-negative} 
Wigner function, controlling
the  position and momentum distribution of each particle. 
Nevertheless, there exist 
measurement schemes in which an analog of Bell's inequality is
violated \cite{Gisin91,GisinPeres92}  for 
such a state---as for any pure entangled state.

The realization and
measurement of   the EPR translational correlations   of {\em 
material particles} appears to
be very difficult. There have been suggestions to start with entangled light
fields and to transfer their quantum state into the state of trapped ions in
optical cavities \cite{Parkins} or of vibrating mirrors 
\cite{ZhangBraunstein}.
We have proposed to realize translational EPR states by taking advantage of interatom
correlations in a dissociating diatom \cite{opa01}.
More recently, we have considered dipole-dipole coupled cold atoms in
an optical lattice as a source of translational EPR states \cite{PRL03}.

In order to generate the translational EPR entanglement between 
interacting material particles, 
one must be able to   accomplish several challenging tasks: 
(a)   switch on and off the entangling interaction;    (b) confine their motion
to single   dimension, and (c) infer and verify the dynamical variables of   
particle 2   {\em at the time of measurement\/} of particle 1. The latter
requirement   is particularly hard for free particles, since by the time we
complete the prediction   for particle 2, its position will have changed. In  
\cite{opa01} we suggested to    overcome these hurdles by transforming the
wavefunction of flying    (ionized) atoms emerging from diatom
dissociation  by an electrostatic/magnetic lens
onto the image   plane, where its  position corresponds to what it was at
the    time of the diatom dissociation.  In  \cite{PRL03} we have
proposed a  solution based on the following steps: 
(i) controlling the    diatom formation and  
dissociation in an optical lattice by switching on and off a laser-induced
dipole-dipole interaction; (ii)    controlling the motion
and effective masses of   the atoms and the diatom by changing the intensities
of the lattice fields.    In this article we discuss our proposal in more
detail and elaborate on its principles.

Our aim here is to demonstrate
the feasibility of preparing a momentum- and position-entangled state
of atom pairs in optical lattices, which would be a variant of the 
original EPR state, owing to lattice diffraction.
In Sec. \ref{Sec-specif}  we specify the
physical system under study.  In Sec. \ref{Sec-singles} the basic properties of
single-atom states  in optical lattices are discussed.
In Sec.
\ref{Sec-binding} we discuss the binding effect of the dipole-dipole
interaction. Sec. \ref{Sec-prep} deals with the preparation of EPR states
by manipulation of the effective masses of the atoms.  In Sec. \ref{Sec-measur}
we discuss experimental demonstration possibilities of measuring the 
EPR
Sec. \ref{Sec-conclusion} is devoted to conclusions.

\begin{figure}[t!]  
\centerline{\epsfig{figure=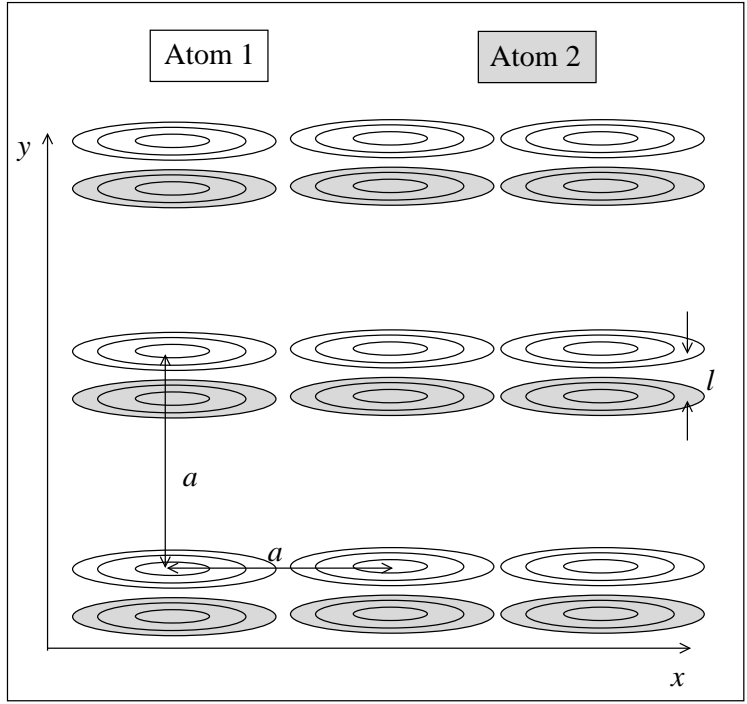,width=0.6\linewidth}}  
\caption{Proposed  
scheme of overlapping optical lattices used to prepare
the
translational EPR state. The lattices are
displaced from each other in the $y$ direction by $l$.
They are sparsely  
occupied by two kinds of atoms. Each of the  
two kinds of atoms feels a different lattice; the shaded regions depict the  
energy minima (potential wells) of the lattices.   
\label{fig-lattice}  
}   
\end{figure}  

\begin{figure}[t!]  
\centerline{\epsfig{figure=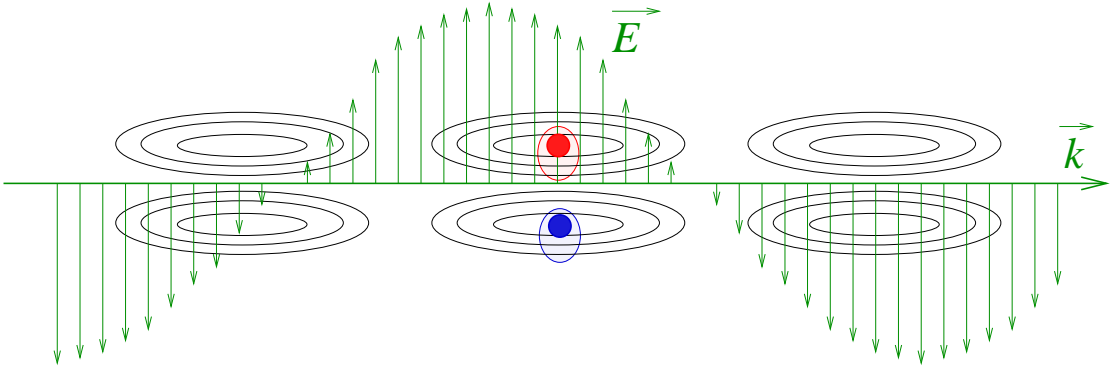,width=0.9\linewidth}}  
\caption{Scheme of the LIDDI interaction: a traveling laser field propagating
in the direction $x$ 
along which the atoms are weakly confined. The electric field
vector is  in the $xy$ plane. The field induces dipole moments in the
 atoms, thereby causing
the interatomic interaction.
\label{f-lattice3}  
}   
\end{figure}  

\section{System description}
\label{Sec-specif}

Let us assume two overlapping optical lattices with the same lattice constant  
$a$, as in Fig. \ref{fig-lattice}.    The lattices are very sparsely occupied
by two kinds of atoms, each kind   interacting with only one of the two
lattices.  This can be realized, e.g.,  by   assuming two different internal
(say, hyperfine) states of the atoms   \cite{Brennen}. In both 
lattices, the potentials
are strongly confining in the $y$   and $z$ directions (realized by strong
laser   fields), whereas in the $x$ direction the lattice potential 
is only
moderately to weakly   confining. Thus, the motion of each particle 
is restricted
to the   $x$ direction. In each direction we assume that only the lowest
vibrational   energy band is occupied.    Initially, the potential minima of
the  lattices are displaced    from each other by an amount $l \ll a$ in the
$y$ direction.  
An auxiliary laser produces a laser-induced
dipole-dipole interaction (LIDDI) between the atom pairs.   
It is linearly
polarized in the $y$ direction, traveling in the $x$ direction and has
a
wavelength $\lambda_{\rm   C}$, moderately
detuned from an atomic transition that {\em differs} from   the
transition used to trap the
atoms in the lattice. In the case of two atoms with identical
polarizabilities in the geometry of Fig. \ref{f-lattice3}, the
interatomic LDDI potential induced by a 
linearly polarized laser is of the form \cite{Thirun}
\begin{eqnarray}  
V_{\rm dd}=-V_{\rm C} F_{\theta}(k R),
\end{eqnarray}
where
\begin{eqnarray}
 F_{\theta}(k R) = \cos \left(   
 k R \cos \theta \right) 
  \left\{ (2\! -\! 3\cos^2 \theta) \left[ \frac{\cos k R}  
 {(k R)^3}\!  + \! \frac{\sin k R}{(k R)^2} \right]\!   
 + \! \cos^2 \theta \frac{\cos k R}{k R}  
 \right\},  
\end{eqnarray}
and 
\begin{eqnarray}  
 V_{\rm C} = \frac{\alpha^2  
 k^3 I_{\rm C}}{4 \pi \epsilon_0^2 c}.
\end{eqnarray}  
Here the wavenumber is $k=2\pi/\lambda_{\rm C}$, $I_{\rm C}$  is   
the coupling laser intensity,  and the atomic dynamic  
polarizability $\alpha$ is 
\begin{eqnarray}
\alpha = \frac{2 \omega_A |\mu|^2}{\hbar (\omega_A^2-\omega^2)},  
\end{eqnarray}
$\mu$ being the   dipole moment element, $\omega_A$ the atomic transition
frequency, and   $\omega = kc$. The position-dependent part $F_{\theta}(k R)$
is a   function of $R$, the distance between the atoms, and $\theta$, the angle
between the interatomic axis and    the wavevector of   the coupling laser.   
Since $l\ll 2a$, $V_{\rm dd}(R)$   has a pronounced minimum for atoms located
at   the nearest sites, $R\simeq l$, where 
\begin{eqnarray}
V_{\rm dd}(R)\simeq   -\frac{ V_{\rm
C}}{4 \pi^3}
\left( \frac{\lambda_{\rm C}}{l}\right) ^3  .   
\end{eqnarray}

\begin{figure}[htb]  
\centerline{\epsfig{figure=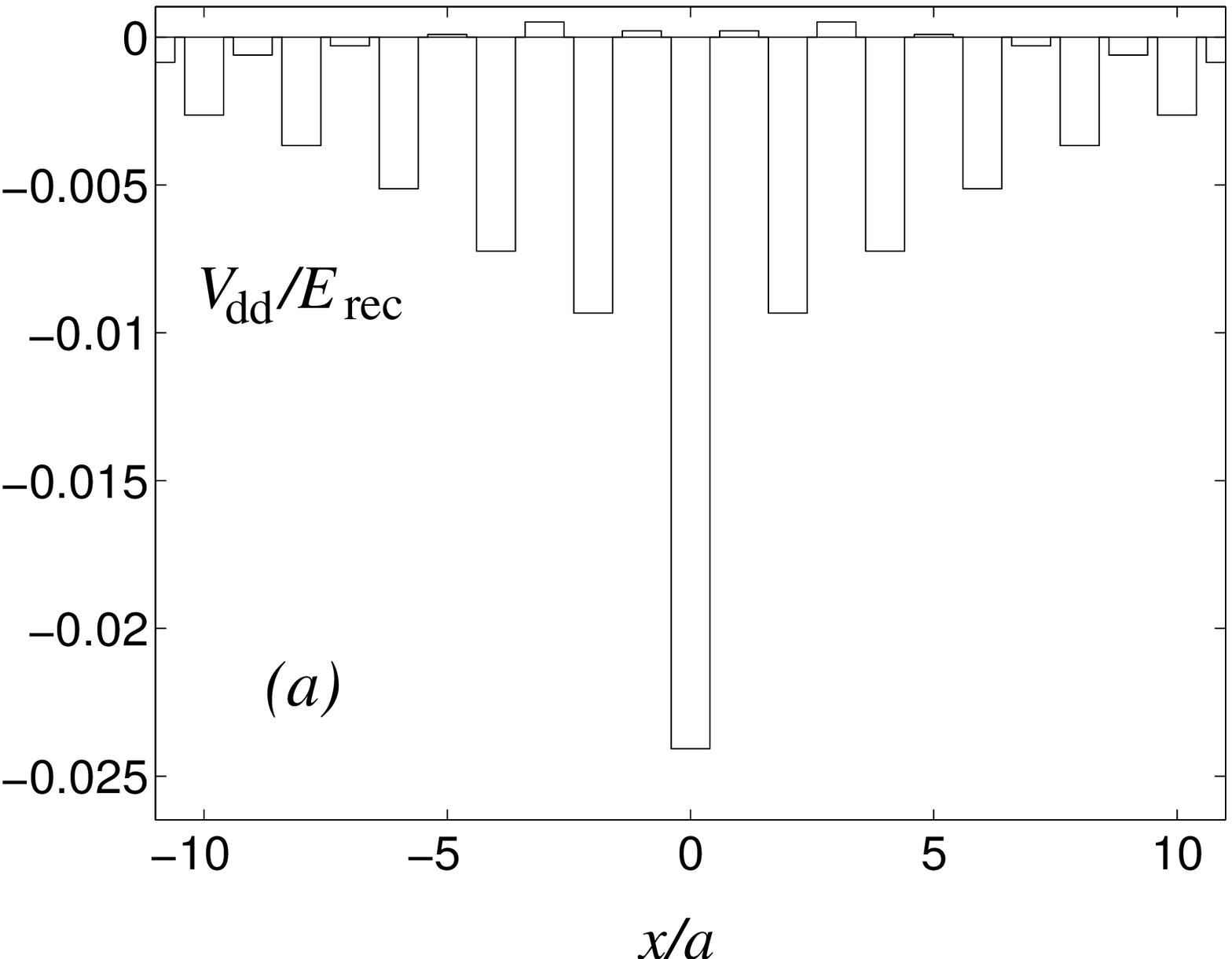,width=0.33\linewidth}
\epsfig{figure=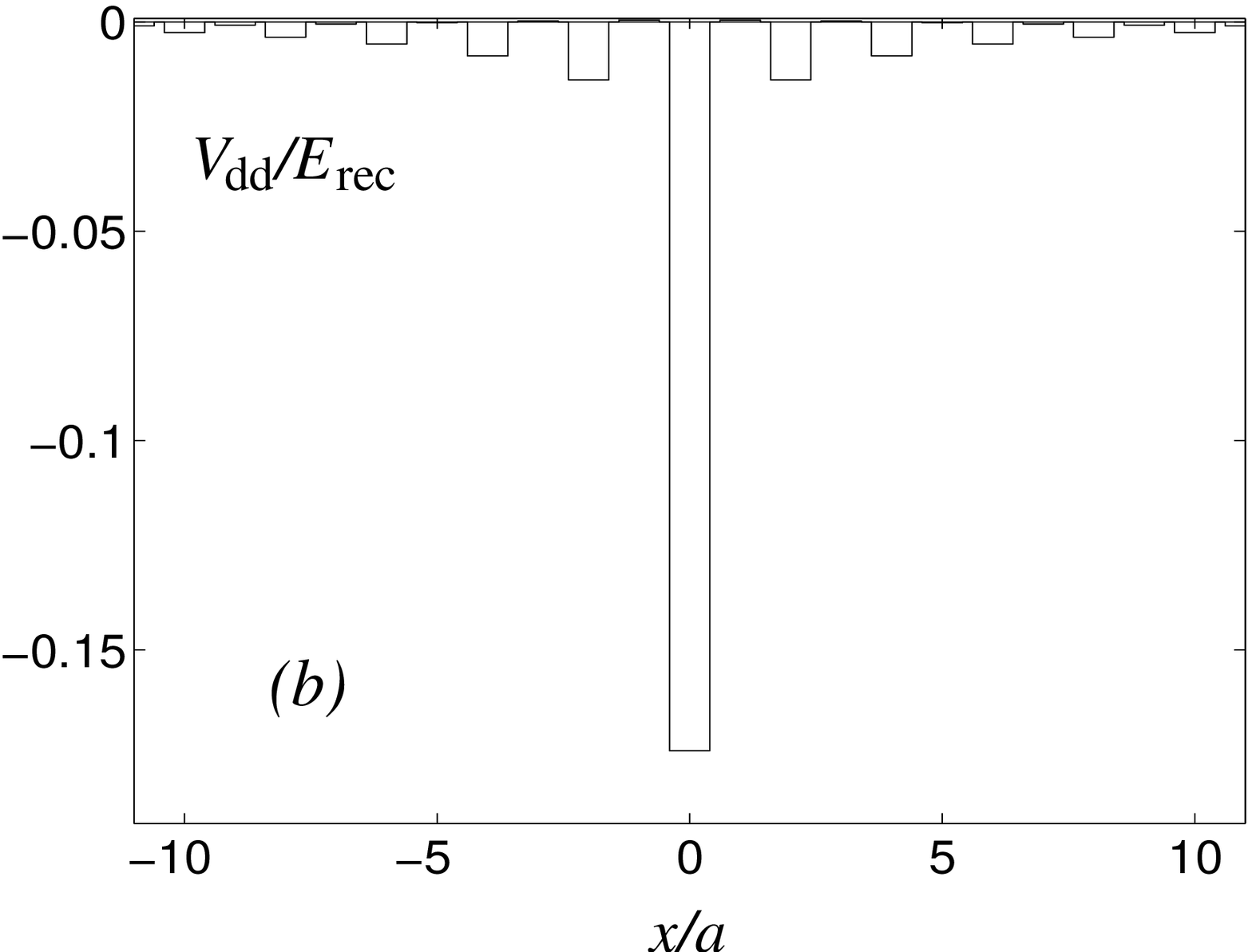,width=0.33\linewidth}
\epsfig{figure=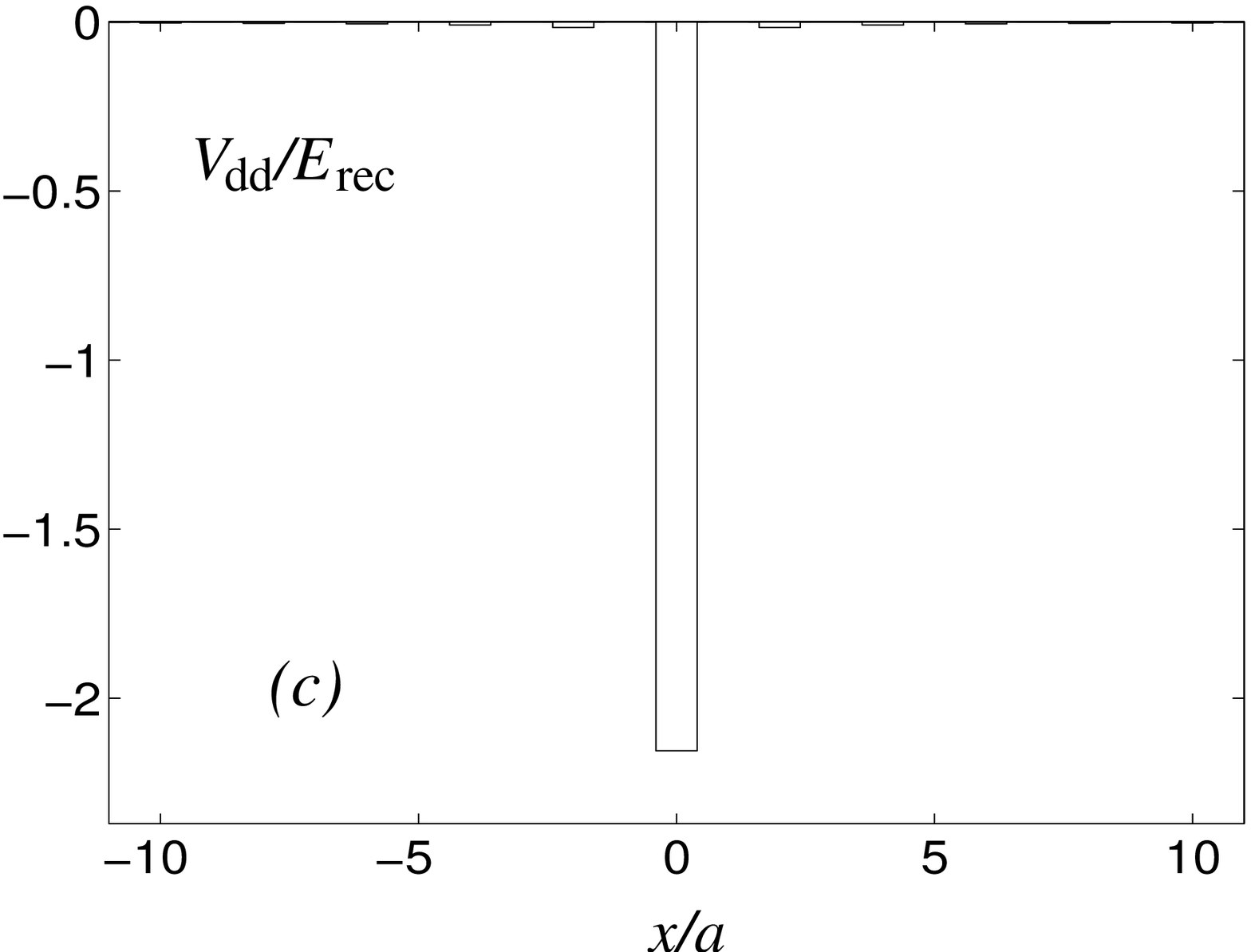,width=0.33\linewidth}
}  
\caption{  
LIDDI potential as a function of the position of atom 2, given that 
atom 1 occupies
site 0, for different separations of the two lattices: (a) $l=200$~nm, (b)
$l=100$~nm, (c) $l=40$~nm. The other parameters are specified in Sec. \ref{Sec-prep}.
\label{f-ddInt}  
}   
\end{figure}

\begin{figure}[htb]  
\centerline{\epsfig{figure=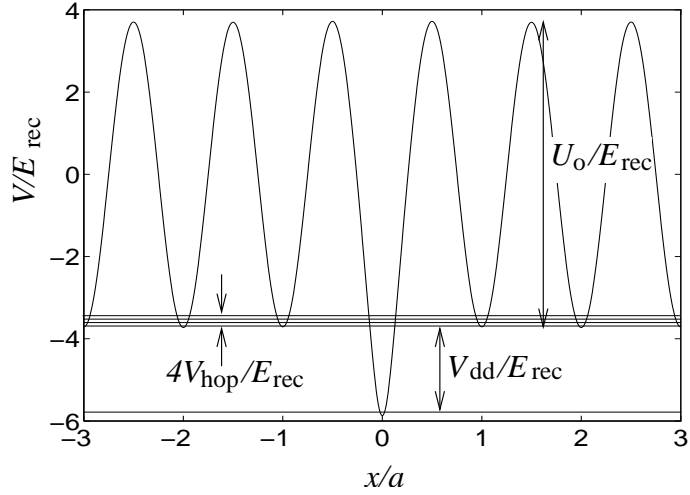,width=0.55\linewidth}}  
\caption{  
Position dependence of the
potential energy of atom 2, given that atom 1 is located at site 0.
The horizontal lines denote the lowest band of energies corresponding 
to uncorrelated
atoms (bandwidth $4V_{\rm hop}$) and the band of diatom  energies
$V_{\rm dd}$ (below the band of uncorrelated atoms).
\label{f-potent1}  
}   
\end{figure}

The LIDDI energy as a function of $l$ and the relative position of the atoms is
shown in Figs. \ref{f-ddInt} and \ref{f-potent1}. 
Under the above  assumptions, we can treat the system as consisting    of pairs
of ``tubes'', either   empty or occupied, that are oriented along $x$.   Only
atoms within adjacent tubes are appreciably attracted to each   other   along
$y$, due to the LIDDI.

\section{Single-atom states in the Wannier basis}
\label{Sec-singles}

Let us focus on the subensemble of tube-pairs in which each tube  is  
 occupied
by exactly one atom. In the 1D optical lattice, the single-atom Hamiltonian is    
\begin{eqnarray}
\hat H_{\rm lat}
= \frac{U_0}{2} \cos \left( \frac{2\pi x}{a} \right) + 
\frac{\hat p_x^2}{2m}.
\label{eq-Hlat2} 
\end{eqnarray}
Here  $m$   is the
atomic mass, $\hat p_x$ is the momentum operator,
$U_0$ is the maximum potential energy due to the
interaction of the atomic dipole with the laser field,
\begin{eqnarray} 
 U_0 = \frac{4 |\mu_L|^2}{\epsilon_0 \hbar c \delta_L} I_L,
\end{eqnarray}
where $\mu_L$ is the dipole matrix element of the lattice transition, 
$\delta _L$ is the detuning of the lattice field from this 
transition, and $I_L$ is the
intensity of the lattice field.
The Hamiltonian (\ref{eq-Hlat2}) describes a quantum
pendulum. The eigenfunctions of the corresponding Schr\"{o}dinger 
equation are
the Mathieu functions \cite{Slater}. The eigenvalues form bands, whose
spectrum depends on the ratio of $U_0$ to the recoil energy,  
\begin{eqnarray}
E_{\rm rec}
= \frac{2\pi^2 \hbar^2}{m   \lambda_{\rm L}^2},
\end{eqnarray}    
so that one can distinguish between strongly binding ($U_0 \gg E_{\rm rec}$)
and weakly binding ($U_0 \sim E_{\rm rec}$) potentials.
We assume that the atoms are cooled down to the lowest
energy band of the lattice, in the absence of LIDDI. 

The state of each
atom is then conveniently described in terms of Wannier functions $|\chi_j\rangle$
\cite{Wannier} that are localized    at lattice sites labeled by index $j$. The
Wannier functions are superpositions of the delocalized Bloch
eigenfunctions $|\phi_k\rangle$ of
the same band,
\begin{eqnarray}
 |\chi_j\rangle = \frac{1}{\sqrt{N}} \sum_{k} \exp \left( - i k x_j \right)
 |\phi_k \rangle , 
\end{eqnarray}
where $N$ is the number of lattice sites, and $x_j$ is the position of the $j$th
site.
Since the Wannier functions are not eigenfunctions of the
single-particle Hamiltonian, an atom initially  prepared in a Wannier
state that is localized at one site, will subsequently tunnel 
to the neighboring sites.
Nevertheless, if the tunneling rate is sufficiently slow, the single-particle
Hamiltonian \r{eq-Hlat2} in the Wannier basis has a relatively simple form: 
\begin{eqnarray}
 H_{\rm lat} \approx \left(
 \begin{array}{cccccc}
 \dots & \dots & \dots & \dots & \dots & \dots \\
 \dots & H_0 & V_{\rm hop} & 0 & 0 & \dots \\
 \dots &  V_{\rm hop}   &   H_0   &   V_{\rm hop}   &  0   &  \dots   \\
 \dots &  0  &   V_{\rm hop}   &   H_0   &   V_{\rm hop}   &  \dots   \\
 \dots &  0  &   0  &    V_{\rm hop}  &   H_0   &  \dots   \\
 \dots & \dots & \dots & \dots & \dots & \dots \\
 \end{array}
 \right) .
 \label{eq-Hlat}
\end{eqnarray}
Here the diagonal elements $H_0$ are equal to  the energy at the 
center of the band, and
only two sets of off-diagonal elements, expressing hopping between 
the neighboring sites,
are non-negligible:
\begin{eqnarray} 
H_0=\langle \chi_j|\hat H_{\rm   lat}|\chi_{j}\rangle,\ \ 
 V_{\rm hop}=\langle \chi_j|\hat H_{\rm   lat}|\chi_{j+1}\rangle,
 \label{eq-Vhop} 
\end{eqnarray}
The hopping rate is related
to the energy bandwidth of the lowest lattice band $V_{B}$ by $V_{B}\approx 4  
|V_{\rm hop}|$ (for exact expressions see \cite{Slater}).

For a moderately deep lattice potential
($U_0 \lesssim 15 E_{\rm rec}$), the quantum-pendulum
Schr\"{o}dinger equation yields the
approximate formulae for $V_{\rm hop}$ and the single-atom effective
mass:
\begin{eqnarray}
V_{\rm hop} & \approx & \frac{1}{4}E_{\rm rec} \exp \left( -
0.26  \frac{ U_0}{E_{\rm rec}} \right) ,
 \\
m_{\rm eff} &=& \frac{2 \hbar^2}{a^2 V_B} 
\approx \frac{\hbar^2}{2 a^2 |V_{\rm hop}|}
\label{15}
\end{eqnarray} 
The Wannier functions of the lowest band 
can be approximated by Gaussians:
\begin{eqnarray}
 \langle x|\chi_j \rangle \approx 
 \langle x|\psi_j^{\rm Gauss} \rangle =
 \frac{1}{\sqrt{2\pi\sqrt{\sigma_G}}}
 \exp \left( - \frac{(x-x_j)^2}{4 \sigma_G^2} \right) ,
\end{eqnarray}
where 
\begin{eqnarray}
 \sigma_G^2 = \frac{\lambda_L^2}{4\pi^2} \sqrt{\frac{E_{\rm rec}}{2U_0}} .
 \label{eq-disp}  
\end{eqnarray} 
This approximation is relatively accurate for $U_0 \gtrsim 6 E_{\rm rec}$
(see the inset in Fig. \ref{figconprobx}) with fidelity 
$|\langle \psi_j^{\rm Gauss}|\chi_j\rangle |^2 > 98$\%. 
Note, however, that the
Gaussian approximation is not suitable for calculating the hopping potential
(\ref{eq-Vhop}) since this quantity is very sensitive to the
non-Gaussian tails of the Wannier 
wavefunctions.


\section{Diatom binding and translational EPR states}
\label{Sec-binding}

\begin{figure}[htb]  
\centerline{\epsfig{figure=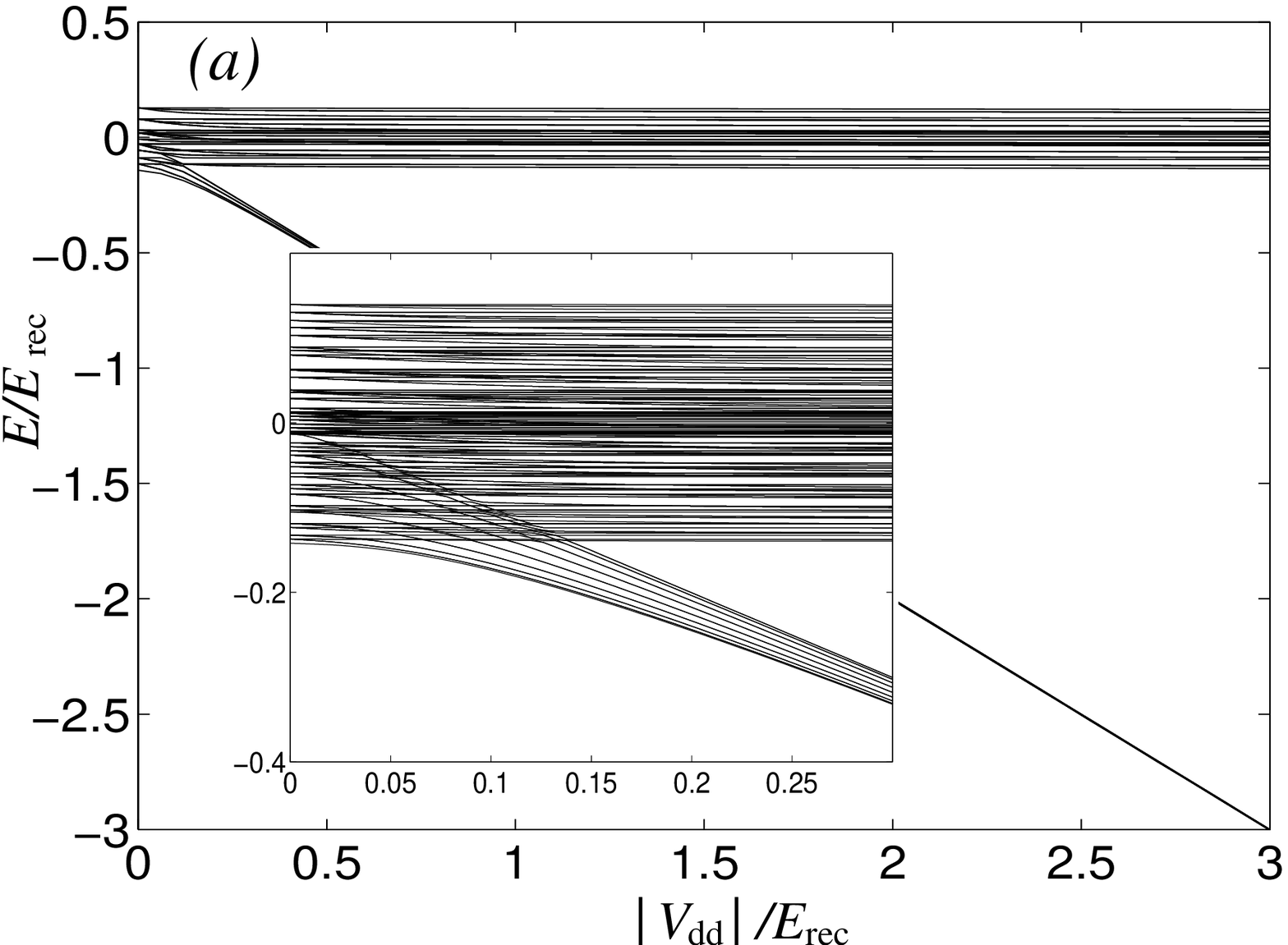,width=0.45\linewidth}
\epsfig{figure=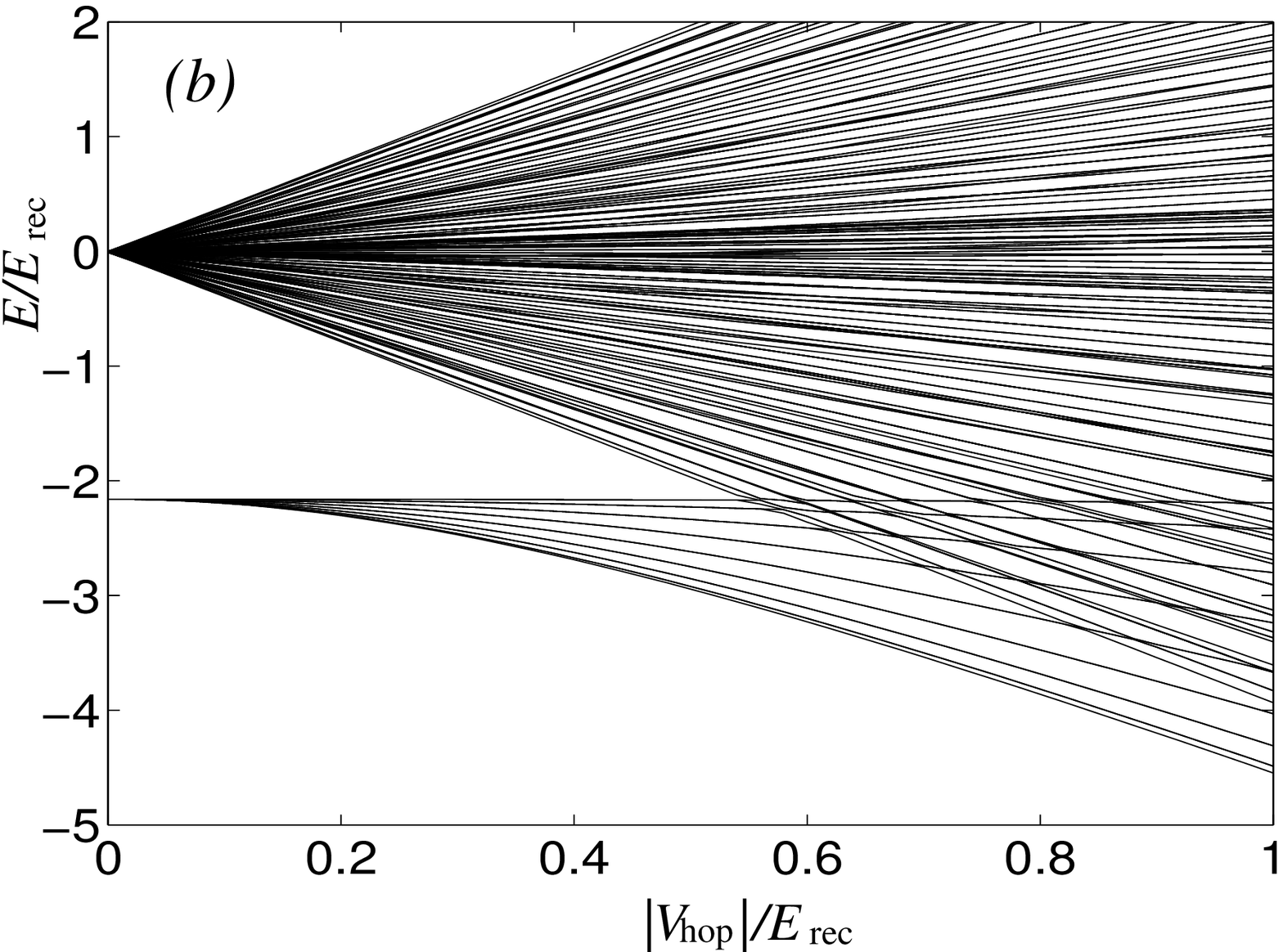,width=0.45\linewidth}}  
\caption{  
Eigenvalues of the two-atom Hamiltonian as a function of $(a)$
the dipole-dipole coupling 
$V_{\rm dd}$ (for a constant hopping potential 
$|V_{\rm hop}| = 0.0355~E_{\rm rec}$),
and $(b)$ the hopping potential 
$V_{\rm hop}$   (for a constant dipole-dipole coupling potential
$V_{\rm dd} = 2.16~E_{\rm rec}$).
\label{f-enVdd}  
}   
\end{figure}

Let us now assume that two neighboring tubes in Fig. \ref{f-lattice3}
are occupied by one
atom each and the LIDDI is turned on. If the tubes are close to each other ($l \ll
\lambda_{\rm C}$, as in Fig.  \ref{f-ddInt}), the interaction Hamiltonian in
the Wannier basis has nonzero elements only for atoms residing at the 
nearest
sites,
\begin{eqnarray}
 \hat H_{\rm int} \approx V_{\rm dd}\sum_{j}
 |\chi_{j}^{(1)}\rangle|\chi_{j}^{(2)}\rangle
 \langle \chi_{j}^{(1)}|\langle \chi_{j}^{(2)}|,
\end{eqnarray} 
and the total two-atom Hamiltonian is 
\begin{eqnarray}
 \hat{H}^{\rm (2at)} = \hat H_{\rm lat}^{(1)}\otimes \hat 1 ^{(2)}
+ \hat 1 ^{(1)}\otimes \hat H_{\rm lat}^{(2)} + \hat H_{\rm int} .
\label{eq-totalham}
\end{eqnarray}
This Hamiltonian has been diagonalized numerically, and its 
eigenvalues 
are shown in Fig. \ref{f-enVdd}
as a function of the hopping and binding potential strength.
One can see that for a
sufficiently large ratio $|V_{\rm dd}|/|V_{\rm hop}|$ a band of diatomic states
is split off the band of independent atoms, towards lower energies.

\begin{figure}[t!]  
\centerline{\epsfig{figure=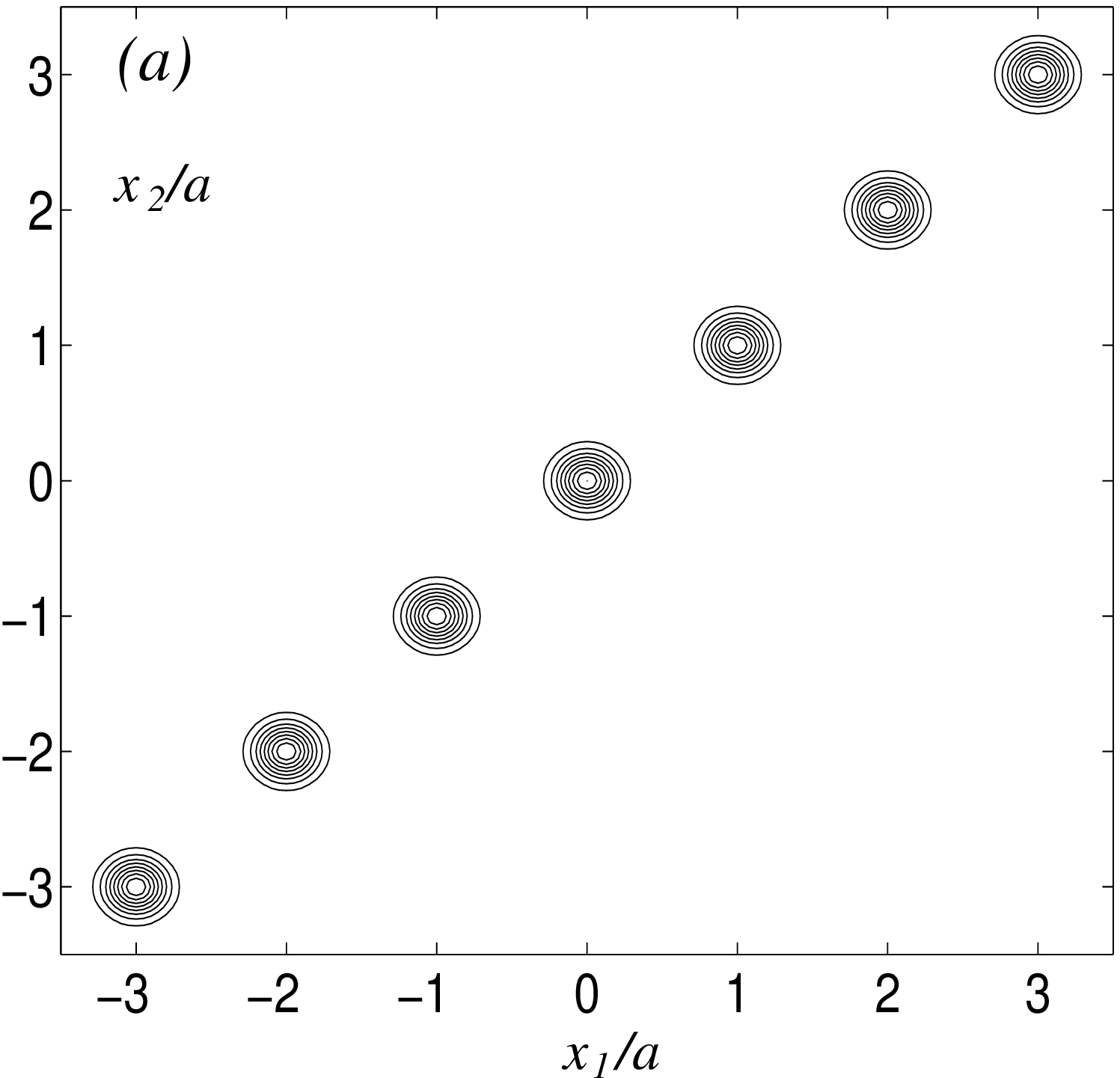,width=0.45\linewidth}
\epsfig{figure=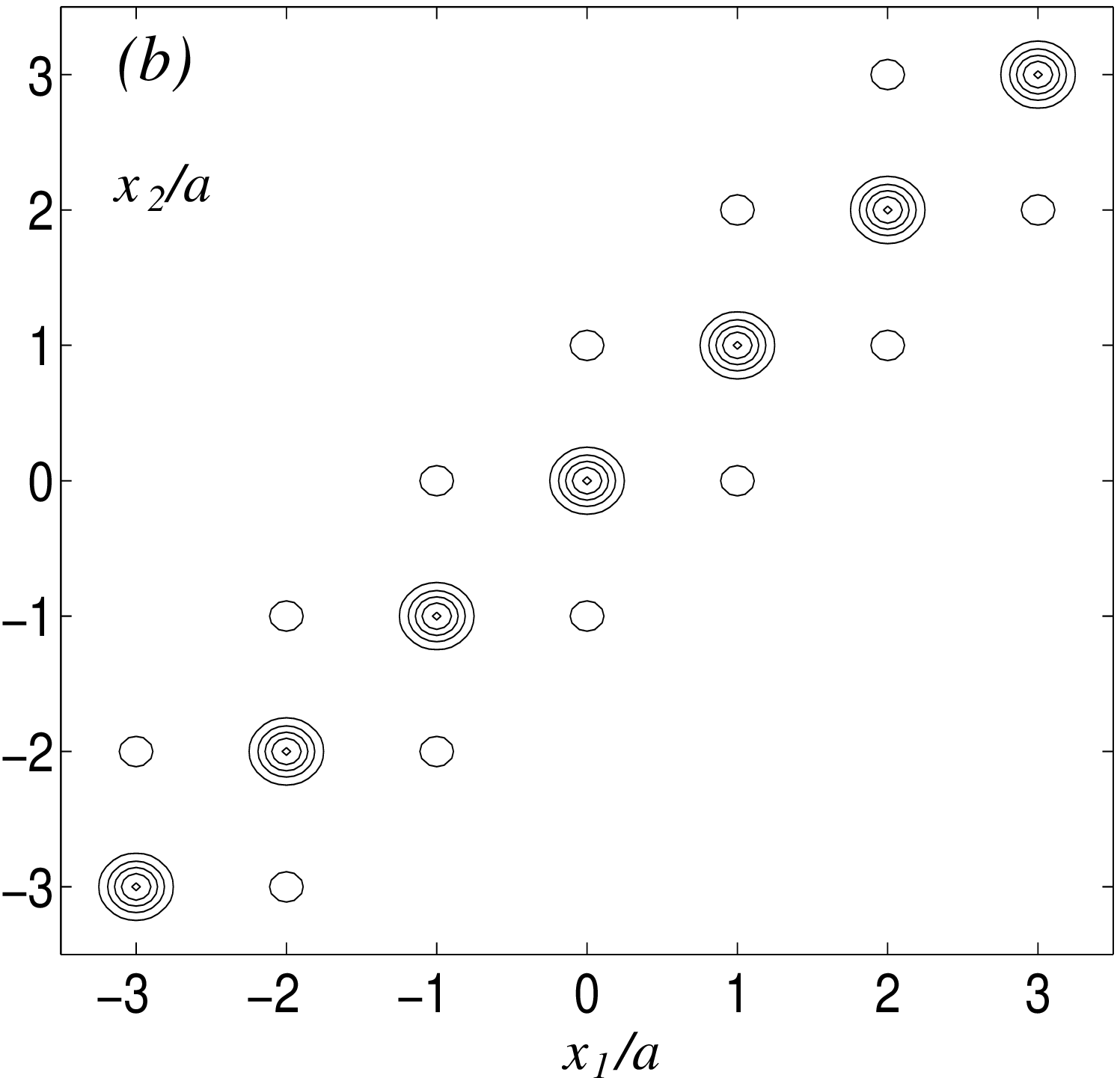,width=0.45\linewidth} } 
\caption{  
Joint probability distribution of the positions of two atoms   
in the ground state of Hamiltonian (\ref{eq-totalham}) for $|V_{\rm hop}|
= 0.0355 E_{\rm rec}$: $(a)$ $|V_{\rm dd}| = 1.0 E_{\rm rec}$, $(b)$
$|V_{\rm dd}| = 0.10 E_{\rm rec}$.
\label{fig-xx} 
}   
\end{figure}

\begin{figure}[t!]  
\centerline{\epsfig{figure=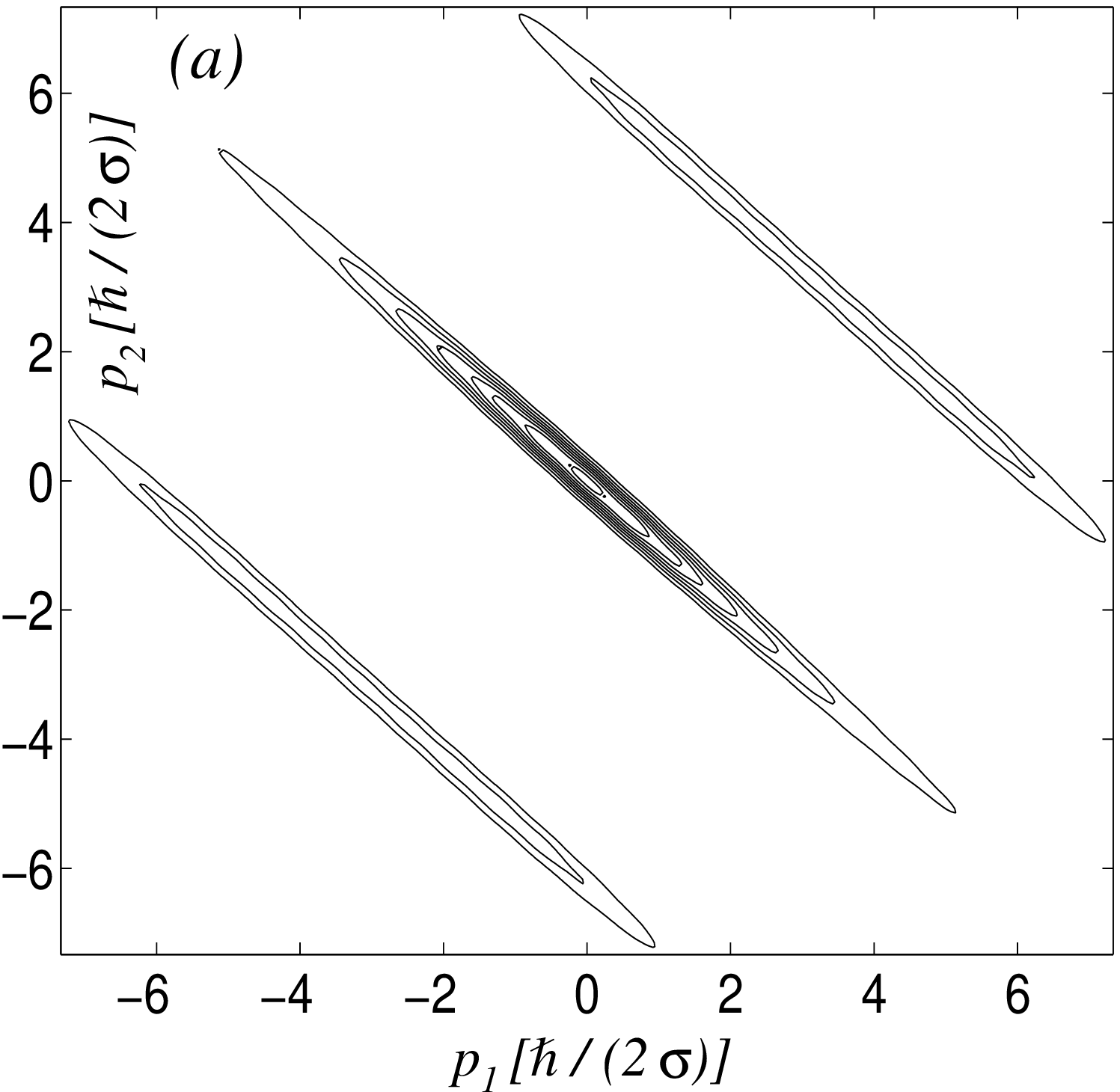,width=0.45\linewidth}
\epsfig{figure=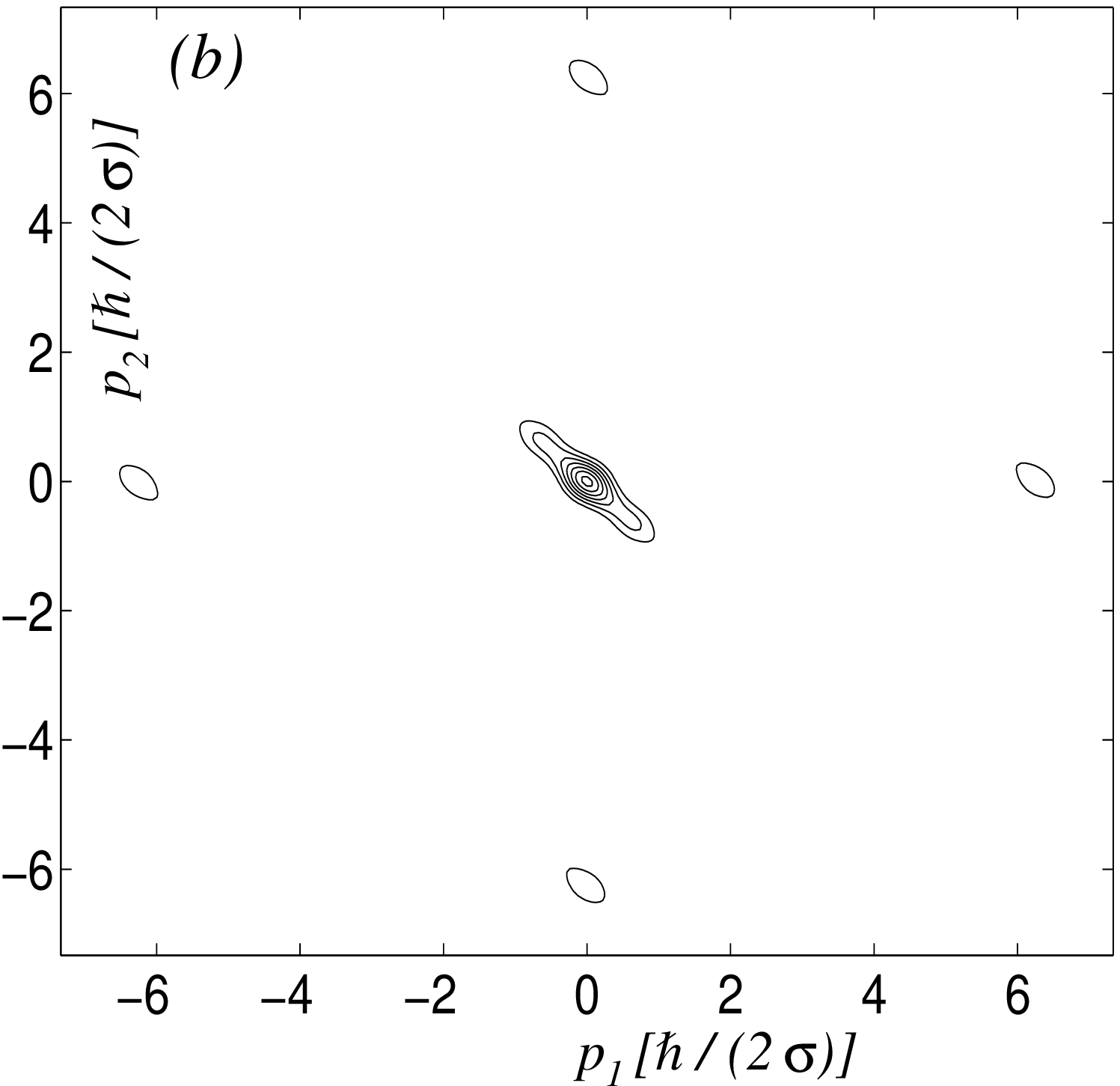,width=0.45\linewidth}  }
\caption{  
Joint probability distribution of the momenta of two atoms   
in the ground state of Hamiltonian (\ref{eq-totalham}) for
$|V_{\rm hop}|
= 0.0355 E_{\rm rec}$: $(a)$ $|V_{\rm dd}| = 1.0 E_{\rm rec}$, $(b)$
$|V_{\rm dd}| = 0.10 E_{\rm rec}$.
\label{fig-pp} 
}   
\end{figure}

For  a strong  LIDDI binding, $|V_{\rm hop}| \ll |V_{\rm  
dd}|$, the ground state of the Hamiltonian  \r{eq-totalham}
corresponds to a tightly bound diatom which can be  approximated by    
\begin{eqnarray}
 |\psi_0\rangle \approx \frac{1}{\sqrt{N}}  
 \sum_{j}|\chi_{j}^{(1)}\rangle|\chi_{j}^{(2)}\rangle.
\end{eqnarray}   
This is a highly correlated state:  when particle 1 is found at the    $j$th
site of lattice 1,    then particle 2 is found   at the   $j$th site of lattice
2, with   position dispersion given by the half-width $\sigma$ of the atomic  
Wannier   function in the lowest band, $\sigma \approx \sigma_G$ (Fig.
\ref{fig-xx}a). 
The Fourier transform of this wave function yields its
momentum representation. The corresponding momentum probability 
distribution exhibits anti-correlation similarly to the EPR states (\ref{EPRstateDelta}) or
(\ref{EPRstateGauss}), but it reflects the lattice periodicity (Fig. \ref{fig-pp}a). 
In momentum space, the state occupies a
region of half-width $\hbar/(2 \sigma)$, and the probability distribution
has narrow ridges along $p_2 = - p_1$. The width of the
ridges is inversely proportional to the lattice size,  $\Delta
p_+\sim \hbar/(Na)$, and they are shifted by $2\pi \hbar /a$ from each
other. 
  
The probability of atoms to
escape their EPR partners ``over the next'' sites  increases with 
the ratio $V_{\rm hop}/V_{\rm dd}$. This leads to an increase of
the position dispersion which can be estimated by first-order perturbation theory:
Let atoms 1 and 2 occupy the $j$th site in the absence of 
$V_{\rm hop}$. With the perturbation $V_{\rm hop}$ on, atom 2 can occupy also
sites $j\pm 1$, which have energies $|V_{\rm dd}|$ above the 
unperturbed state, with
the probability $\approx |V_{\rm
hop}|^2/|V_{\rm dd}|^2$. This contributes to an increase in the
diatomic    separation dispersion, 
\begin{eqnarray}
 \Delta x_{-}^2 \approx \sigma^2 + 2 a^2 \left(   
 \frac{V_{\rm hop}}{V_{\rm dd}} \right) ^2,
\end{eqnarray} 
resulting in the joint probability distribution of the atomic 
positions and momenta as a function of
$V_{\rm hop}/V_{\rm dd}$, shown in Fig.
\ref{fig-xx}b and Fig.
\ref{fig-pp}b, respectively. 

The  states of the tightly bound diatom form a separate band whose  
bandwidth is 
\begin{eqnarray}
 V_{B}^{\rm (2 at)} \approx 4 |V_{\rm hop}^{\rm (2 at)}|,
 \label{bandwidthdiat}
\end{eqnarray}  
below the lowest atomic  
vibrational band.   
The {\em diatomic\/}   
hopping potential $V_{\rm hop}^{\rm (2  
at)}$ can be estimated by   
assuming that the two atoms  
consecutively  hop to their neighboring sites, i.e., the change
\begin{eqnarray}   
 |\chi_{j}^{(1)}\rangle|\chi_{j}^{(2)}\rangle \to   
 |\chi_{j+1}^{(1)}\rangle|\chi_{j+1}^{(2)}\rangle
\end{eqnarray}
is realized   
either via  
\begin{eqnarray}
 |\chi_{j}^{(1)}\rangle|\chi_{j}^{(2)}\rangle \to   
 |\chi_{j+1}^{(1)}\rangle|\chi_{j}^{(2)}\rangle \to   
 |\chi_{j+1}^{(1)}\rangle|\chi_{j+1}^{(2)}\rangle, 
\end{eqnarray}
or via  
\begin{eqnarray}
 |\chi_{j}^{(1)}\rangle|\chi_{j}^{(2)}\rangle \to   
 |\chi_{j}^{(1)}\rangle|\chi_{j+1}^{(2)}\rangle \to   
 |\chi_{j+1}^{(1)}\rangle|\chi_{j+1}^{(2)}\rangle .
\end{eqnarray}   
By adiabatic elimination of the higher-energy intermediate states,
one obtains 
\begin{eqnarray}
 V_{\rm hop}^{\rm (2 at)} \approx 2 \frac{|V_{\rm hop}|^2}{ V_{\rm dd}}.
\end{eqnarray}   

All the states of the diatomic band have correlated positions. 
However, the momenta are not anti-correlated in all these states in 
the same way as in the diatomic ground state.
To realize strong momentum  
anti-correlations, 
we have to prepare a state that predominantly originates from the 
bottom of the diatomic band.
If we work with thermal states this means that
the temperature of the system must satisfy   
\begin{eqnarray}
k_B T \ll V_{B}^{\rm (2 at)}.  
\end{eqnarray}
Near the bottom of the band, the diatomic dynamics can be described 
by means of the 2-atom effective mass
given by
\begin{eqnarray}
 m_{\rm eff}^{\rm (2 at)} = \frac{2\hbar^2}{V_{B}^{\rm (2 at)}a^2}  
 \approx \frac{\hbar^2 |V_{\rm dd}|}{4V_{\rm hop}^2 a^2}.  
\end{eqnarray}
The thermal (kinetic) energy of the diatom is then related to the 
degree of momentum anti-correlation through
the sum-momentum spread $\Delta p_{+} = p_{x1}+p_{x2}$,
\begin{eqnarray}
 \Delta p_{+}^2 \approx  k_{B} T
 {2 m_{\rm eff}^{\rm (2 at)}}
\approx \frac{\hbar^2 |V_{\rm dd}|}{4V_{\rm hop}^2 a^2}  
 k_B T.  
\end{eqnarray}

To determine how ``strong'' the EPR effect is, we compare the product of  
the half-widths of the position and momentum peaks in the tightly  
bound diatom state  
with the Heisenberg uncertainty limit through the parameter \cite{opa01,PRL03}:  
\begin{eqnarray}   
s = \frac{\hbar}{2\Delta x_{-}\Delta 
 p_{+}}. 
 \label{sparameter}   
\end{eqnarray}   
A value of $s$ higher than 1 indicates the occurrence of the EPR effect; the  
higher the value of $s$, the stronger the effect. 

Strictly speaking, 
for  
the multi-peak momentum distribution, one should use a more general
uncertainty relation, as  
discussed, e.g., in \cite{Uffink}, that distinguishes the  
uncertainty of multiple
narrow peaks from that of a single broad peak.  
However, even the simple  
half-width of the peaks  
is a useful measure of the  
EPR effect.  
In order to maximize $s$, we must adhere to the  
trade-off between reducing either $\Delta x_{-}$ , by decreasing  
$|V_{\rm hop}/V_{\rm dd}|$, or $\Delta p_{+}$, by  
increasing $|V_{\rm hop}/V_{\rm dd}|$. The optimum value of $s$ generally   
depends on  
the temperature of the diatom, as detailed below.

\section{EPR state preparation}
\label{Sec-prep}  
 
Cooling down the diatomic system  to prepare the    EPR state is a 
non-trivial
task. We suggest   a ``cooling''  procedure which takes advantage of the
difference between the single-atom and the 
diatom  bandwidths, and
of the possibility  to change the light-induced potentials. 
The key is first to cool down individual atoms and then separate
the unpaired
atoms from the diatoms. 
The scheme consists of three steps:  

(i)   We first switch on only
an external,    shallow, harmonic potential in the $x$ direction (all
other potentials being off), and cool
the     $x$-motion    of the atoms down to its ground state. The width $\sigma_E$ of the ground state should be several times
the    lattice constant; it is related to the desired momentum anti-correlation
by    $\sigma_E \approx \hbar/(\sqrt{2} \Delta p_{+})$.  The temperature 
necessary to
achieve this must be    
\begin{eqnarray}   
T\ll \hbar^2/(4m k_B \sigma_E^2). 
\end{eqnarray}   

(ii)    A weak
lattice potential in the $x$-direction is then slowly switched     on, so that
the state becomes  
\begin{eqnarray}   
(\sum_{j}\alpha_j    |\chi_{j}^{(1)}\rangle )
(\sum_{l}\alpha_l|\chi_{l}^{(2)}\rangle ) =     \sum_{j}\alpha_j^2
|\chi_{j}^{(1)}\rangle |\chi_{j}^{(2)}\rangle + \sum_{j\neq    l}\alpha_j
\alpha_l |\chi_{j}^{(1)}\rangle  |\chi_{l}^{(2)}\rangle, 
 \label{33}   
\end{eqnarray}   
where the   
coefficients 
\begin{eqnarray}   
\alpha_j \sim \exp [-(j-j_0)^2 a^2/(4 \sigma_E^2)] 
\end{eqnarray}   
are   
Gaussians localized  around the minimum of the external potential.    

\begin{figure}[t!]   
\centerline{\epsfig{figure=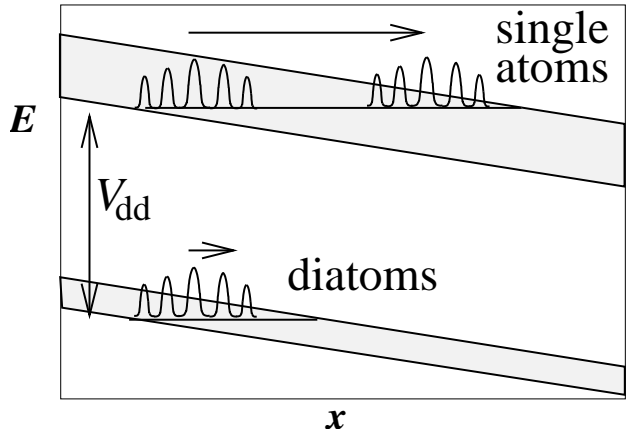,width=0.5\linewidth}}   
\caption{
Separating single (unpaired) atoms from diatoms: an external 
repulsive potential causes both the single atoms and the diatoms to 
move on a surface of
constant energy. The diatoms hit the top of the energy band after being
displaced by a much shorter length than the unpaired atoms.
\label{pasy}
}    
\end{figure}

 (iii)
We    switch on the LIDDI and change the external potential,    from 
an attractive
well to     a repulsive linear potential, acting to remove the particles from
the lattice.  The    two parts of the wavefunction \r{33} will 
respond differently:
The   motion of the  paired atoms (corresponding to 
$\sum_{j}\alpha_j^2 |\chi_{j}^{(1)}\rangle |\chi_{j}^{(2)}\rangle$) 
will remain in the vicinity of the
initial position because of their narrow energy band (Fig.~\ref{pasy}). 
Single (unpaired) atoms, whose bandwidth is substantially larger, will travel a much longer distance before hitting the top of the energy band.
Thus, after a properly chosen time, the
unpaired atoms could be removed from the $x$-region of interest, which is $\approx
V_{\rm hop}/V^{(2)}_{\rm hop}$ longer for single atoms than for diatoms [see
Eq. ~(\ref{bandwidthdiat})]. 

Provided the time $t_c$ is short enough for the system to remain
near the bottom of these two
bands, the dynamics can be interpreted in terms of the appropriate effective 
masses: the
``heavy'' diatoms with mass move much slower than the ``light'' 
single atoms with mass (\ref{15}).
Thus, after changing the sign of the external potential,
the unpaired atoms will be
ejected out of  
the lattice  and separated from the diatoms as glumes from   
grains.

\begin{figure}[t!]   
\centerline{\epsfig{figure=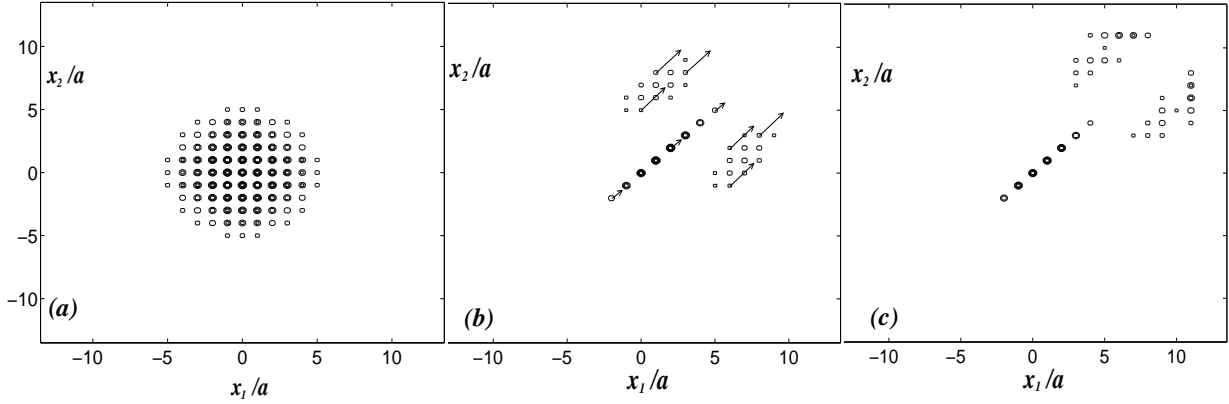,width=0.99\linewidth}}   
\caption{
Simulation of the EPR state preparation in an optical lattice with 25 
sites, at three consecutive times:
(a)~Initially ($t=0$),  the atoms are cooled down to the external harmonic potential
ground state, whereas the LIDDI is off. (b)~At $t=1.4\times 10^{-4}$ s LIDDI and the
repulsive linear potential (with the slope 0.04 $E_{\rm rec}$ per lattice site) 
are on, whereas the harmonic potential is off. 
The diatoms are moving
through the lattice {\em very} slowly in comparison to the single 
atoms.
(c)~At $t=2.16\times 10^{-4}$ s single atoms are ejected out of the lattice and the     
diatoms are separated out.      
\label{celek}
}    
\end{figure}

This effect is illustrated by the numerical simulation in  
Fig.~\ref{celek} for two lithium atoms
in two lattices with   $\lambda_{\rm L}=$ 323 nm
(corresponding to the transition 2s--3p) and a   dipole-dipole coupling field
of $\lambda_{\rm C}=$ 670.8 nm (transition   2s--2p). The dipole moment
element of the lattice transition is $1.26\times 10^{-30}$ Cm, while 
the LIDDI coupling dipole element is
$2.7\times 10^{-29}$ Cm. From these values we get 
the recoil energy $E_{\rm rec}=1.85\times 10^{-28}$J. 
The lattice and LIDDI field intensities are $I_{\rm L} =$ 0.186
W/cm$^2$ and $I_{\rm   C} =$ 0.023 W/cm$^2$. The corresponding field 
detunings are
$\delta_{\rm L} = 50   \gamma_{\rm L}$, $\delta_{\rm C} = 100 \gamma_{\rm C}$,
the respective decay rates   being $\gamma_{\rm L} = 1.2 \times 10^6$ s$^{-1}$, and
$\gamma_{\rm C} =   3.7 \times 10^7$ s$^{-1}$. The two lattices are displaced
by $l=$ 40~nm.   {F}rom these values we get the lattice potential $U_0 = 3.93
E_{\rm rec}$,   the LIDDI potential of the nearest atoms $V_{\rm dd} =
-0.5   E_{\rm rec}$, and the hopping potential $V_{\rm hop}=-0.09 E_{\rm
rec}$.   The two-particle hopping potential is then $V_{\rm hop}^{\rm
(2at)}\approx   -0.0324 E_{\rm rec}$.   The correlated pairs are prepared by
first cooling   independent atoms in an external harmonic potential with the
ground-state   half-width of $\sigma_E = 5a$ (frequency of 1 kHz $\sim
30$~nK). When the linear external potential is switched on, the atoms start
moving in the direction of decreasing potential energy.  
Figure~\ref{celek}, which captures the situation at three consecutive
times, shows that unpaired atoms (off-diagonal peaks) are displaced by a much longer distance
than the diatoms (diagonal peaks).

The paired atoms remaining in the lattice are then in the state    $\sim \exp
[-(j-j_0)^2 a^2/(2 \sigma_0^2)]    |\chi_{j}^{(1)}\rangle
|\chi_{j}^{(2)}\rangle$ wherein positions and momenta are correlated
   with the uncertainties
$\Delta x_+\approx \sigma_E/\sqrt{2}$ and $\Delta
p_{+}\approx\hbar/\Delta x_+$, respectively.    At higher temperatures the atoms are not
cooled to the ground state of the    external potential and the momentum
anti-correlation has the spread  
\begin{eqnarray} 
\Delta p_{+}   
\approx\hbar/\{ \sqrt{2}\sigma_E  \tanh [\hbar^2/(2\sigma_E^2 mk_B T)]\}.  
\end{eqnarray} 
The   
parameter $s$ of Eq. (\ref{sparameter}) can then be estimated as   
\begin{eqnarray}   
 s \approx \frac{\sigma_E}{\sqrt{2} \sigma}   
 \tanh \left[ \frac{1}{\pi^2} \left( \frac{a}{\sigma_E} \right)   
 ^2 \frac{E_{\rm rec}}{k_B T} \right].   
\end{eqnarray}   
This relation enables us to select the optimal external harmonic potential   
(specified here by $\sigma_E$)     such that the parameter $s$ is maximized,   
for a given temperature $T$.   

The small effective mass of unpaired atoms allows us to cool them   
to temperatures higher than     that
corresponding to the bottom of the diatomic band. The    price is, 
however,
that most of the atoms are discarded and only     a small    fraction 
of $\sim    a/\sigma_E$ will remain in the
bound diatom state.    The    different behavior of the paired and unpaired
atoms in a periodic     potential is a sparse-lattice analogy of the
transition from
Mott-insulator to a superfluid    state in the fully occupied lattice, recently
observed in Ref. \cite{Greiner}.

The two-particle   joint position distribution of the 
ground state is    a chain of peaks of   half-width $\sigma$ 
separated by $a$ 
that are located along the line   $x_2=x_1$ (Fig. \ref{figconprobx}). The 
corresponding joint momentum distribution spreads over an area of   half-width 
$\hbar/(2\sigma)$ and consists of ridges in the direction $p_2=-p_1$,
that are separated by $2\pi \hbar/a$, and have   the 
half-width $\pi \hbar/(Na)$ for a lattice of $N$ sites (Fig.    \ref{figconprobp}).


  
\section{Measurements}
\label{Sec-measur} 
 
\begin{figure}[b!]  
\centerline{\epsfig{figure=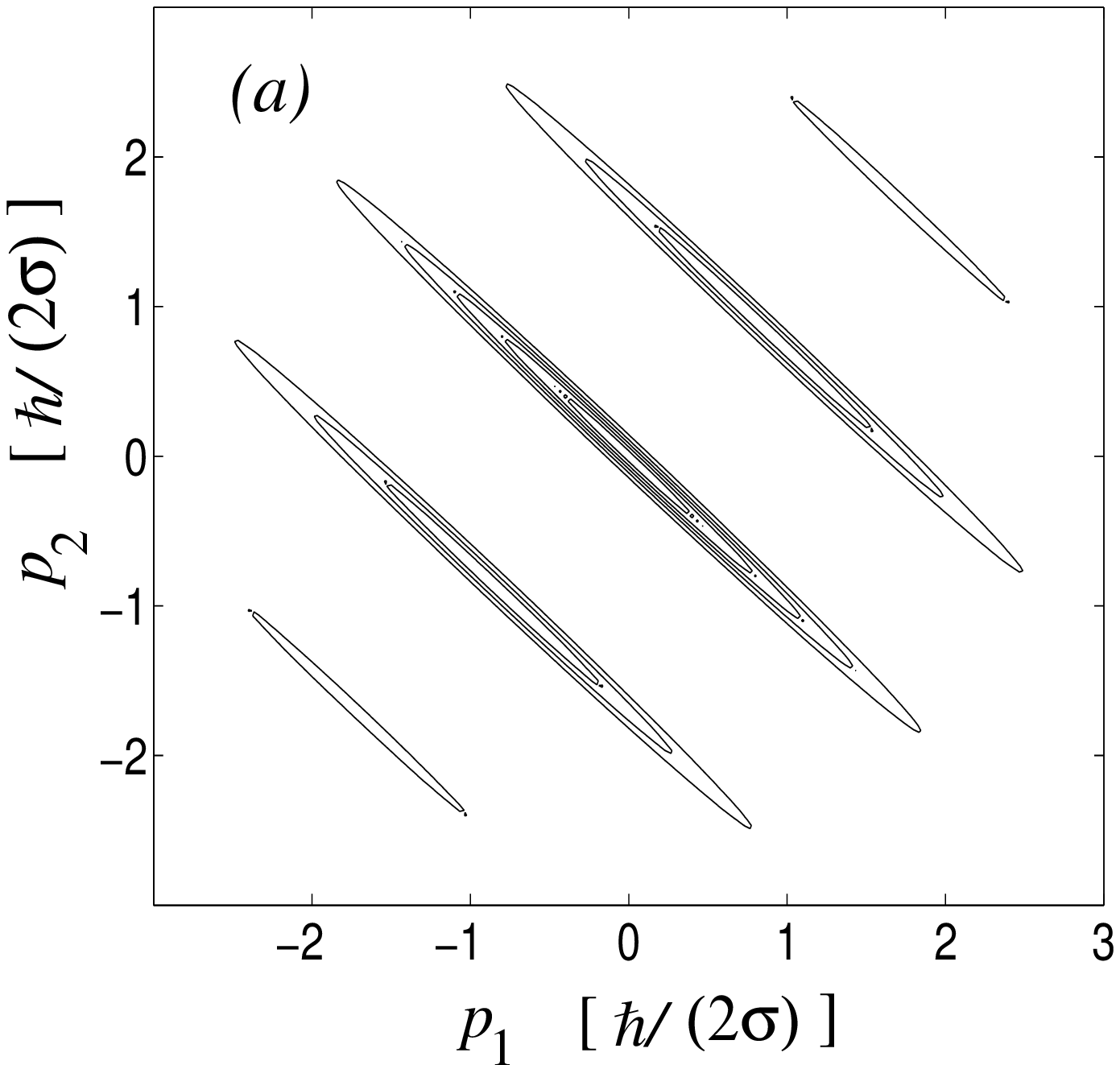,width=0.35\linewidth}
\epsfig{figure=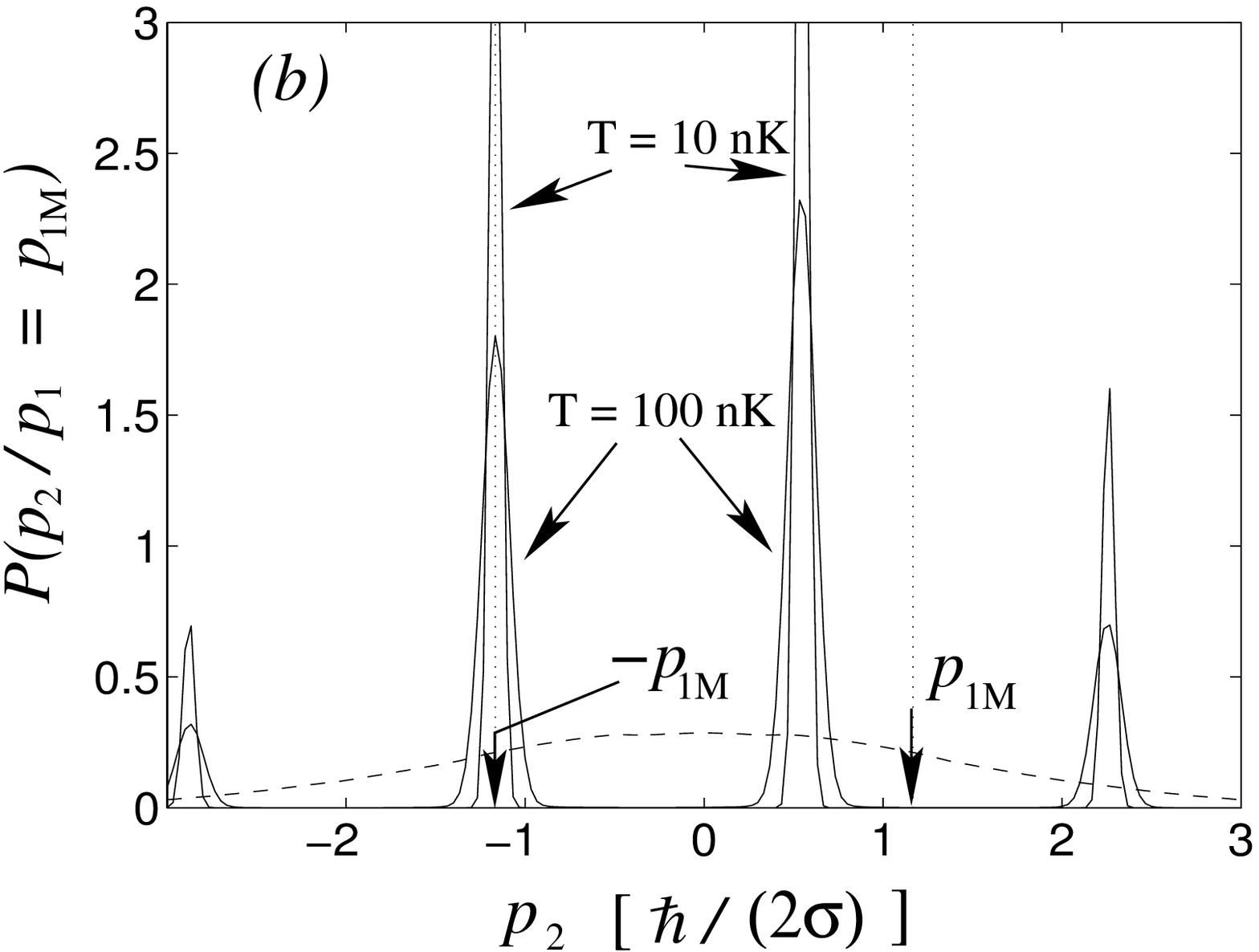,width=0.45\linewidth}}  
\caption{  
$(a)$ Joint   probability distribution of the atomic momenta in  the 
aforementioned state
with  $T=100$~nK.  
$(b)$
Conditional probability of the momentum of atom 2, given that the momentum of atom 1
has been   measured  (the measured
value $p_1=p_{1{\rm M}}$ is indicated by an arrow)for lithium diatoms prepared as in the text.     The dashed line  
corresponds to the marginal probability distribution of momentum $p_2$  
irrespective of the momentum of atom 1 at the temperature $T=100$ nK.   The
half-width of each peak is equal   to $1/s$ of Eq. (\ref{sparameter}).    
\label{figconprobp}  
}   
\end{figure}

After preparing the system in the EPR state, how can one can test its properties  
experimentally? To this end we may increase the lattice potential $U_{0}$,  
switch off the field inducing the LIDDI, and separate the two lattices by  
changing the laser-beam angles. By increasing $U_0$, the atoms lose their  
hopping ability and their quantum state is  
``frozen'' with a large  effective mass: the  
bandwidth $V_B$ decreases exponentially with $U_0$ and the effective   
mass increases exponentially,  
so that the atoms become too ``heavy'' to move. One has then   
enough time to perform measurements on each atom:

\begin{figure}[b!]  
\centerline{\epsfig{figure=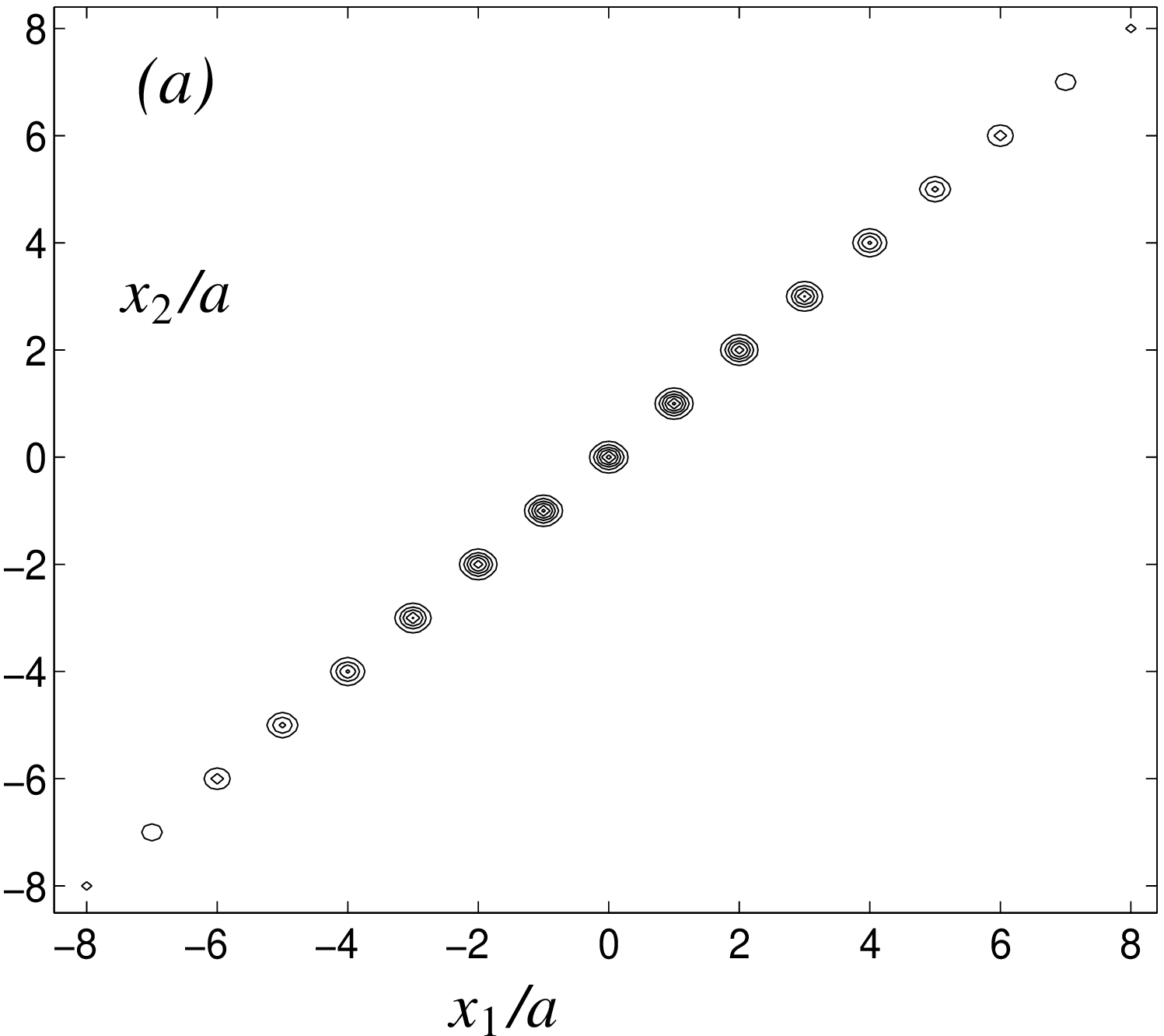,width=0.4\linewidth}
\epsfig{figure=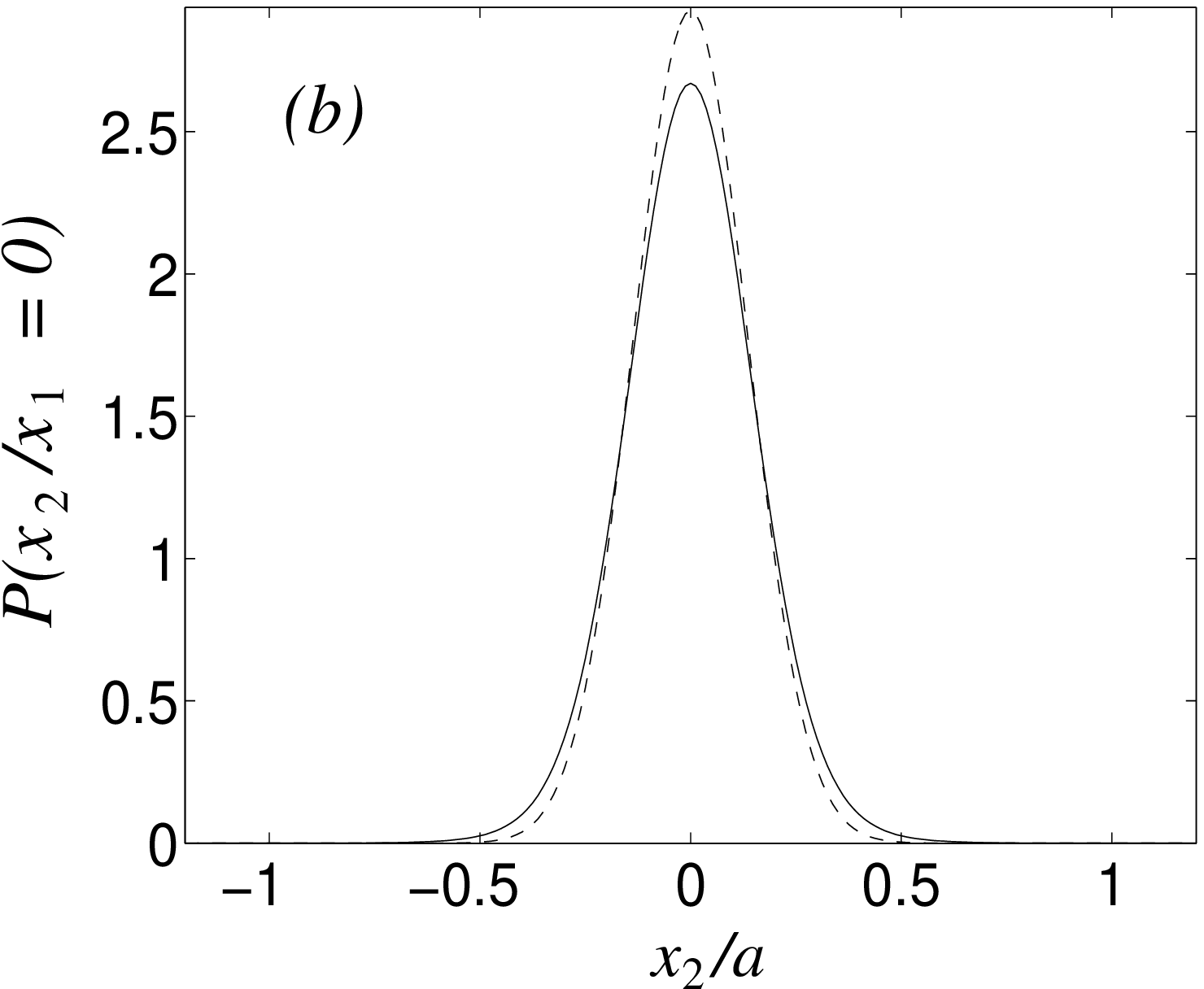,width=0.45\linewidth}}  
\caption{  
$(a)$ Joint probability distribution of the positions of two lithium atoms   
in adjacent  
optical lattices, prepared in a diatom state as specified in the   
text, using  
the ground state of the external harmonic potential 
with half-width $\sigma_E=6a$  
and temperature of 10~nK.  
$(b)$  
Position probability of atom 2 in the state above, conditional on   
atom 1  
being measured at site 0 (full line).   
Dashed line: Gaussian approximation of the  
Wannier function with the half-width $\sigma = 0.14 a$.   
\label{figconprobx}  
}   
\end{figure}  
    
a) The atomic position can be measured by detecting its resonance   
fluorescence.  
After finding the site occupied by atom 1, one can  
infer the position of atom 2. If this inference is confirmed in a large  
ensemble of measurements, it would suggest that there is an ``element of  
reality'' \cite{EPR} corresponding to the position of particle 2.   
An example of the conditional probability of position of particle 2 after
measuring the position of particle 1 is given in Fig. \ref{figconprobx}. 

b) The 
momentum can be measured by switching off the   
$x$-lattice potential of the  
atom (thus bringing it back to its ``normal'' mass $m$). The distance  
traversed by the atom during a fixed time is proportional to its  
momentum.   
An example of the conditional probability of the momentum of particle 
2 given the momentum of particle 1 is shown in Fig. \ref{figconprobp}.

c) One can test the EPR correlations between the  
atomic {\em ensembles} occupying the two lattices, testing large number of pairs  
in a single run.  
The correlations in $x$ and anti-correlations in $p$ would be  
observed by matching the {\em distribution histograms} measured on  
atoms from the two lattices.

\section{Conclusions} 
\label{Sec-conclusion}

We have discussed a scheme which can be
used to prepare a translationally entangled pair of
massive particles in a state analogous to the original EPR state 
\cite{EPR}.   
A novel element of the present scheme is the extension of
the EPR   correlations to    account for lattice-diffraction effects.  
Their momentum and position correlations principally differ from those of free particles [Eqs. 
 \r{EPRstateDelta}, \r{EPRstateGauss}]:   due to the lattice 
periodicity, the {\em position and momentum   distributions have generally a 
multi-peak structure}. 

The realization of the proposed scheme is expected to be based on the 
adaptation of
existing   techniques (optical trapping, cooling, controlled dipole-dipole  
interaction). to the requirements spelled out in Sec. \ref{Sec-prep}
and \ref{Sec-measur}
  The most
important ingredient of the scheme   is the manipulation of the effective mass,
for EPR-pairs  preparation (by separating the   ``light'' unpaired
atoms from the ``heavy'' diatoms)   and for their detection (by ``freezing''
the atoms   in their initial state so that their EPR correlations   are
preserved long enough).   

One may envision 
extensions of the present approach to matter teleportation \cite{opa01}   and
quantum    computation based on continuous   variables
\cite{Braunstein98,Braunstein98N,LloydSlotine98,LloydBraunstein99}.
Such   extensions may involve the coupling of entangled atomic
ensembles in optical lattices by photons carrying quantum information.

\acknowledgments 
We acknowledge the support of ISF, Minerva and the EU Networks QUACS and ATESIT.


\end{document}